\documentclass[AMA,Times1COL]{WileyNJDv5}

\usepackage{graphicx}
\usepackage{siunitx}
\usepackage{bm}
\usepackage{ulem}

\articletype{Original Article}%

\received{Date Month Year}
\revised{Date Month Year}
\accepted{Date Month Year}
\journal{Contributions to Plasma Physics}
\volume{00}
\copyyear{2024}
\startpage{1}

\begin{document}

\newcommand{\titlename}{Electron density structure measurements with scattered intense laser beam}

\title{\titlename}

\author[a]{K. Sakai}
\author[b]{K. Himeno}
\author[c]{S. J. Tanaka}
\author[d]{T. Asai}
\author[o]{T. Minami}
\author[b]{Y. Abe}
\author[b]{F. Nikaido}
\author[b]{K. Kuramoto}
\author[d]{M. Kanasaki}
\author[e]{H. Kiriyama}
\author[e]{A. Kon}
\author[e]{K. Kondo}
\author[e]{N. Nakanii}
\author[f]{W. Y. Woon}
\author[f]{C. M. Chu}
\author[f]{K. T. Wu}
\author[g]{C. S. Jao}
\author[h]{Y. L. Liu}
\author[i]{T. A. Pikuz}
\author[j]{H. Kohri}
\author[k]{A. O. Tokiyasu}
\author[l]{S. Isayama}
\author[m]{H. S. Kumar}
\author[n]{K. Tomita}
\author[e]{Y. Fukuda}
\author[b]{Y. Kuramitsu}

\authormark{SAKAI \textsc{et al.}}
\titlemark{\titlename}

\address[a]{\orgname{National Institute for Fusion Science}, \orgaddress{\state{Toki}, \country{Japan}}}
\address[b]{\orgdiv{Graduate School of Engineering}, \orgname{Osaka University}, \orgaddress{\state{Suita}, \country{Japan}}}
\address[c]{\orgdiv{Department of Physical Sciences}, \orgname{Aoyama Gakuin University}, \orgaddress{\state{Sagamihara}, \country{Japan}}}
\address[d]{\orgdiv{Graduate School of Maritime Sciences}, \orgname{Kobe University}, \orgaddress{\state{Kobe}, \country{Japan}}}
\address[o]{\orgdiv{Graduate School of Information Science and Technology}, \orgname{Osaka University}, \orgaddress{\state{Suita}, \country{Japan}}}
\address[e]{\orgdiv{Kansai Institute for Photon Science}, \orgname{National Institutes for Quantum Science and Technology}, \orgaddress{\state{Kizugawa}, \country{Japan}}}
\address[f]{\orgdiv{Department of Physics}, \orgname{National Central University}, \orgaddress{\state{Taoyuan}, \country{Taiwan}}}
\address[g]{\orgdiv{Department of Physics}, \orgname{National Cheng Kung University}, \orgaddress{\state{Tainan}, \country{Taiwan}}}
\address[h]{\orgdiv{Institute of Space and Plasma Sciences}, \orgname{National Cheng Kung University}, \orgaddress{\state{Tainan}, \country{Taiwan}}}
\address[i]{\orgdiv{Institute for Open and Transdisciplinary Research Initiative}, \orgname{Osaka University}, \orgaddress{\state{Suita}, \country{Japan}}}
\address[j]{\orgdiv{Research Center for Nuclear Physics}, \orgname{Osaka University}, \orgaddress{\state{Ibaraki}, \country{Japan}}}
\address[k]{\orgdiv{Research Center for Accelerator and Radioisotope Science}, \orgname{Tohoku University}, \orgaddress{\state{Sendai}, \country{Japan}}}
\address[l]{\orgdiv{Faculty of Engineering Sciences}, \orgname{Kyushu University}, \orgaddress{\state{Fukuoka}, \country{Japan}}}
\address[m]{\orgdiv{Department of Aerospace Engineering}, \orgname{Tohoku University}, \orgaddress{\state{Sendai}, \country{Japan}}}
\address[n]{\orgdiv{Division of Quantum Science and Engineering}, \orgname{Hokkaido University}, \orgaddress{\state{Sapporo}, \country{Japan}}}

\corres{Kentaro Sakai, National Institute for Fusion Science, 322-6 Oroshicho, Toki, Gifu 509-5292, Japan. \email{sakai.kentaro@nifs.ac.jp}}

\abstract[Abstract]{Short-pulse intense lasers have the potential to model extreme astrophysical environments in laboratories. Although there are diagnostics for energetic electrons and ions resulting from laser-plasma interactions, the diagnostics to measure velocity distribution functions at the interaction region of laser and plasma are limited. We have been developing the diagnostics of the interaction between intense laser and plasma using scattered intense laser. We performed experiments to measure electron density by observing the spatial distributions and ratio of horizontal to vertical polarization components of scattered laser beam using optical imaging. The observed ratio of polarization components is consistent with the drive laser beam indicating the observed light originates from the drive laser. Imaging of the scattered light shows the structure of electron density, the zeros moment of electron velocity distribution function, interacting with the intense laser. We observed the change of structure due to the laser pre-pulse that destroys the target before the arrival of the main pulse.}

\keywords{intense laser, scattering, electron density}

\maketitle

\section{Introduction}

The fundamental physics of space and astrophysical plasmas has been investigated using laser-produced plasmas \cite{takabe21hpl}. So far, model experiments of space and astrophysical phenomena, such as collisionless shocks and magnetic reconnections, have been carried out using supersonic plasma flows (typically 100--\SI{1000}{km/s}) generated by high-power lasers with pulse duration of a few nanoseconds \cite{yamazaki22pre,matsukiyo22pre,sakai22srep,morita22pre,sakai24hedp}. However, it is highly challenging to investigate phenomena driven by relativistic flows, such as relativistic shocks \cite{sironi15ssr} and associated particle acceleration mechanisms \cite{hoshino08apj,iwamoto22apj}, in the current experiments using high-power lasers. We alternatively use short-pulse intense lasers with the pulse duration of \SI{10}{fs}--\SI{1}{ps}, to experimentally simulate space and astrophysical phenomena in laboratories to verify their physical mechanisms \cite{takabe21hpl,kuramitsu15hedp}. 

The recent development of short-pulse intense lasers by chirped pulse amplification technique \cite{strickland85oc} enables us to achieve a fast plasma flow approaching the speed of light and a bright, coherent, and short-pulse light source. The sub-relativistic protons with the velocity of $v= 0.4\text{--}0.5c$, where $c$ is the speed of light, can be generated using the up-to-date intense lasers with the pulse duration of \SI{10}{fs}--\SI{1}{ps} \cite{higginson18ncommun,ziegler24nphys,minami24srep} and using the fast plasma flow, one can model relativistic astrophysical environments in laboratories \cite{kuramitsu15hedp}. Relativistic collissionless shocks and magnetic reconnections have been studied in the laboratory using intense lasers \cite{raymond18pre,law20pre,kuramitsu23pop,campbell24prr}. The intense laser light also can be used for laboratory simulations of extremely bright radiations in the universe. We have been modeling coherent synchrotron emissions in relativistic shocks \cite{kuramitsu11pre,kuramitsu11pop} and pulsar radiations \cite{tanaka20ptep}. 

Although intense lasers have the potential to access the extreme parameter regime, the diagnostics to observe the spatiotemporal evolusion of distribution functions is limited. Most of the diagnostics are time-integrated due to the short timescale (\SI{10}{fs}--\SI{10}{ps}) and require high spatial resolution to resolve the tiny spatial scale (1--\SI{100}{\micro m}). While the time-integrated ex-situ measurements of distribution functions with electron and Thomson parabola spectrometers are the standard diagnostics in the intense laser experiments, there are a few measurements of spatial distribution of plasma parameters or distribution functions. We consider the intense laser itself as a nonlinear probe beam of scattering and have been developing diagnostics to observe nonlinear plasmas generated by the intense laser. Since the scattering of an intense laser beam in a plasma is light scattering by charged particles, the process can be Thomson scattering. In intense laser experiments, imaging diagnostics of Thomson scattering is used to monitor the structure of laser wakefield accelerator \cite{nakanii15prab}.

In the high-power laser experiments, optical and charged particle beam imagings are employed to reveal the spatiotemporal evolution of the electron density and electromagnetic field, respectively \cite{harilal22rmp,schaeffer23rmp}. Since these diagnostics measure line-of-sight integrated value, the quantitative analysis usually assumes cylindrical symmetry. The local density, temperature and velocity of electron and ion are measured with Thomson scattering spectroscopy \cite{froula11}, which is the only local diagnostics in laboratory astrophysics experiments. While plasmas relevant to space and astrophysical phenomena are often nonlinear, non-stationary, non-equilibrium, and unstable, the analysis of Thomson scattering usually assumes linear, stationary, equilibrium and stable plasmas. 
We have been developing diagnostics to measure nonlinear plasmas with Thomson scattering using a linear probe beam that is weak enough not to change the plasma state and the spectral shape is highly different from the linear plasma case \cite{sakai20pop,sakai23pop}. When we use the intense laser as a nonlinear probe, this drastically changes the plasma state within the pulse duration, i.e., generation of nonlinear plasmas. Therefore, the scattering of intense laser is so complex and no theoretical framework to model the scattering process. As a first step to understand the scattering process and to develop diagnostics of intense laser-produced plasmas, we measure the intensity of scattered light that can give a constraint of the scattered spectrum.

In this paper, we report the measurements of spatial distribution of electron density, which is the zeroth moment of electron distribution function with imaging of scattered intense laser beam, as a step toward the measurement of distribution function by spectroscopy. The images show the structures associated with the focusing intense laser beam. We confirmed the images correspond to the scattered intense laser beam using polarimetry. The weak intensity at the focal spot indicates the absence of electrons at the focal spot, which is explained by the combination of pre-pulse and ponderomotive force.

\section{Experiment}\label{sec:exp}

\begin{figure}[tb]
    \centering
    \includegraphics[clip,width=0.7\hsize]{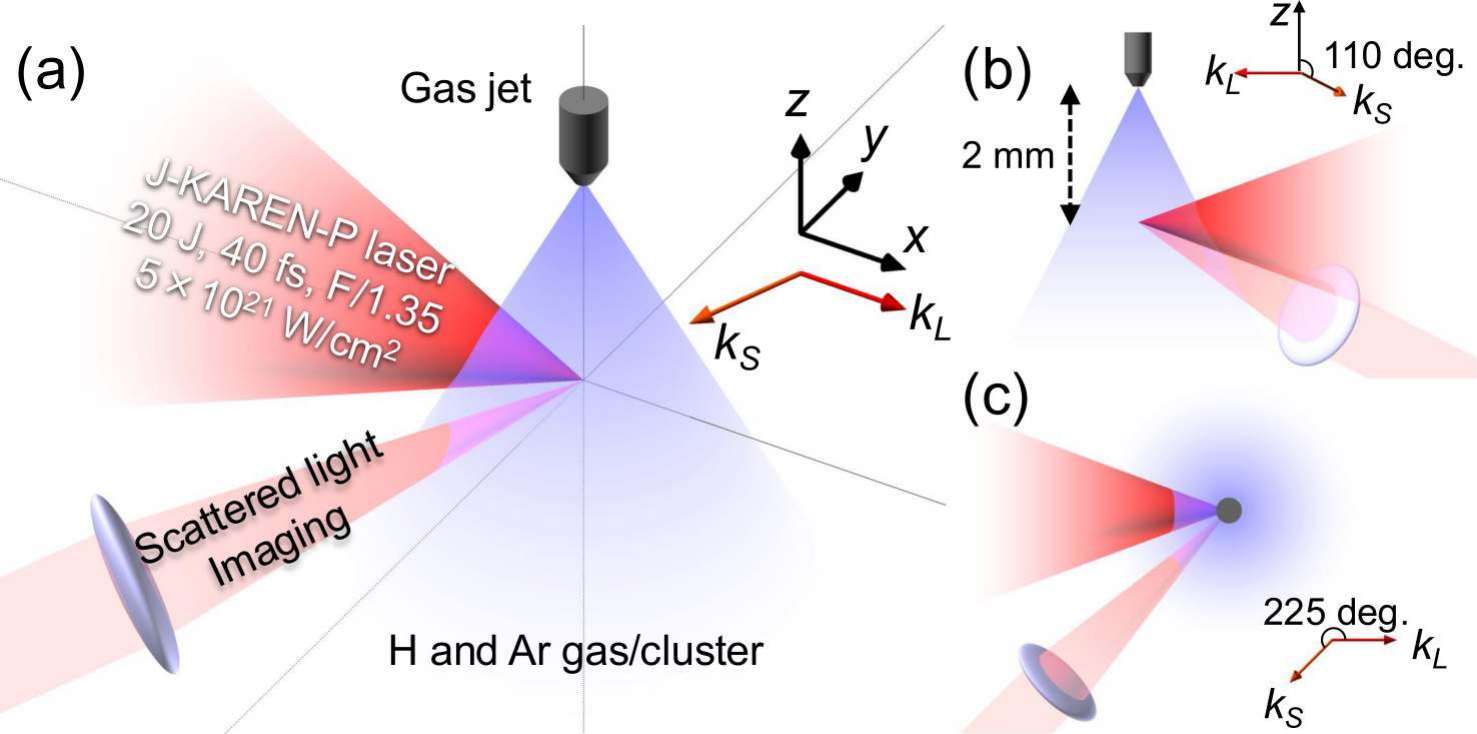}
    \caption{(a) Schematic illustration of the experimental setup. (b) Side view of the setup. (c) Top view of the setup.}
    \label{fig:scheme}
\end{figure}

We performed experiments with the J-KAREN-P laser facility at Kansai Institute for Photon Science, National Institutes for Quantum Science and Technology \cite{kiriyama18ol}. Figure~\ref{fig:scheme}(a) shows the experimental setup with a tight-focusing OAP of F/1.35. The energy, pulse duration, focal spot radius, and peak intensity of the beam are $\varepsilon_L = \SI{20}{J}$, $\tau_L = \SI{40}{fs}$, $r_0 = \SI{1}{\micro m}$, and $I_L = \SI{5e21}{W/cm^{2}}$, respectively. Note that the spot radius is the experimentally measured value and larger than the diffraction limit. The central wavelength of the laser is $\lambda_L = \SI{810}{nm}$. We use the plasma mirror system to control the pre-pulse level \cite{kon22hpl}.
We define the laser propagation direction ($\bm{k_L}$) as the $x$ direction, and the orthogonal horizontal and vertical axes as the $y$ and $z$ directions, respectively. The origin of the orthogonal coordinate is at the focal spot of the laser beam. Hereinafter, this coordinate is referred to as the target chamber coordinate. Above the focal spot at $(0, 0, \SI{2}{mm})$, a high-density cooled gas jet system is located to supply the hydrogen or argon gas-like targets as shown in Fig.~\ref{fig:scheme}(b) \cite{jinno18ppcf,jinno22srep}. By cooling down the gas, the target changes from gas to clusters. 
We operated the gas jet nozzle with the conical angle of \SI{10}{\degree} at room temperature in Sec.~\ref{sec:pol}, with the conical angle of \SI{5}{\degree} at room temperature in Sec.~\ref{sec:pm}, and with the conical angle of \SI{5}{\degree} at \SI{30}{K} in Sec.~\ref{sec:cl}.

We observe the image of the scattered light from the J-KAREN-P laser itself. We use a \SI{50}{mm}-diameter first lens at \SI{250}{mm} away from the origin to collect the scattered light. The scattered light is transported to the detector by silver mirrors and achromatic lenses. The magnification of the imaging system is $\sim 10$. The image is obtained with a CCD camera and provides the spatial distribution of the scattered light. The exposure time of the camera is \SI{10}{ms}, which is much longer than the duration of the main pulse. The imaging system potentially observes both scattered light and self-emission. We adjust the amount of light on the CCD camera using ND filters. We use a bandpass filter to select the measured wavelength of \SI{810}{nm}, which is the laser wavelength. As shown in Figs.~\ref{fig:scheme}(b) and \ref{fig:scheme}(c), the collection direction of the scattered light is $\bm{k_S}$, which is \SI{110}{\degree} from the $z$ direction and \SI{225}{\degree} from the $x$ direction in the $x-y$ plane, thus, the scattered wavevector is written as $\bm{k_S} = k_S (\cos\SI{225}{\degree} \sin\SI{110}{\degree},\sin\SI{225}{\degree}\sin\SI{110}{\degree},\cos\SI{110}{\degree})$. Since the polarization of the laser is in the $y$ direction, the scattered light rarely propagates in the $y$ direction. The collection direction is tilted in the $z$ direction to obtain the scattered light. We use a Wollaston prism with the separation angle of \SI{5}{\degree} to separate the horizontal and vertical polarization components of the scattered light. In Secs.~\ref{sec:pm} and \ref{sec:cl}, we optically rotate the image so that the horizontal axis corresponds to the laser propagation axis.

\subsection{Polarization measurement}\label{sec:pol}

\begin{figure}[tb]
    \centering
    \includegraphics[clip,width=\hsize]{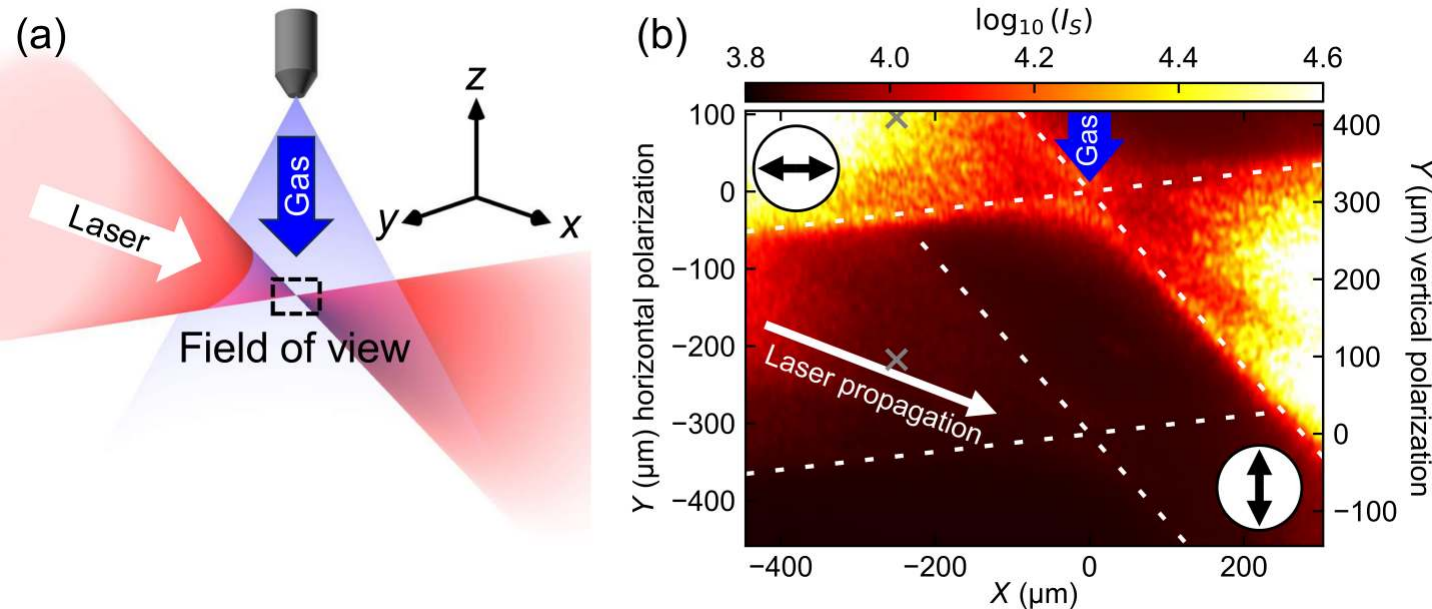}
    \caption{(a) View from the observation direction. (b) 2D image of the scattered light within the dashed square in (a) with the argon gas target. The upper and lower show the horizontal and vertical polarizations on the detector, respectively.}
    \label{fig:imgwp}
\end{figure}

In principle, the observed signal is the mixture of self-emission and scattered light. Because the self-emission originates from random electron motion, the polarization is assumed to be unpolarized. On the other hand, the scattered light maintains the polarization of the incident laser. We observed polarization components to estimate the amount of self-emission relative to scattered light.
Figure~\ref{fig:imgwp} shows a two-dimensional (2D) image of scattered light with the argon gas target, where tiny grain structures of clusters are not observed. Figure~\ref{fig:imgwp}(a) shows the view from the observation direction. The field of view of the observed image in Fig.~\ref{fig:imgwp}(b) is shown in the dashed square.
Ideally, the resolution of the image is given by $\delta = 1.22 l \lambda_L /D \approx \SI{5}{\micro m}$, where $l$ is the distance between the first lens and the origin, $D$ is the diameter of the first lens, and the focal depth is $d = 2\pi \delta^2 / \lambda_L \approx \SI{200}{\micro m}$. We take an average of five images with the same target and laser conditions to improve the signal-to-noise ratio. The vertical ($X$) and horizontal ($Y$) axes in the image are the coordinates projected on the detector. 
Two hyperbolas indicated by the white-dotted curves show the optical path of the ideal Gaussian laser beam with the spot diameter of \SI{2}{\micro m} and F/1.35 projected on the detector coordinate. The upper and lower hyperbolas correspond to the horizontal and vertical polarization components, respectively. Since the intensity is clearly different for two polarization components, the observed signal is polarized.  
The optical path of the focusing drive laser is visualized in the image, which is mostly consistent with the ideal Gaussian laser beam, except for the focal spot size. The observed focal spot size can be attributed to the finite resolution of the imaging system. The laser propagates from the upper left to the lower right. The laser propagation direction is tilted by $\sim \SI{20}{\degree}$. Considering the geometry of measurement, the transformation from the target chamber coordinate to the detector coordinate is given by $(X, Y) = (- x \sin \SI{225}{\degree} + y \cos \SI{225}{\degree}, -x \sin \SI{225}{\degree}\cos \SI{110}{\degree} - y \cos \SI{225}{\degree}\cos \SI{110}{\degree} + z \sin \SI{110}{\degree})$. The angle of the laser propagation axis projected on the detector is $\tan^{-1}(\cos \SI{110}{\degree}) \approx \SI{19}{\degree}$ and is consistent with the observation. The focal spot [$(X,Y) = (0,0)$] is darker than the surroundings. 

The intensity ratio of horizontal to vertical component ($I_h/I_v$) in the current experimental setup is given by
\begin{equation}
\frac{I_h}{I_v} = \frac{\frac{I_s\cos \SI{110}{\degree}}{1+\cos \SI{110}{\degree}} + \frac{I_e}{2}}{\frac{I_s}{1+\cos \SI{110}{\degree}} + \frac{I_e}{2}} \approx \left\{ \begin{array}{ll} \cos\SI{110}{\degree} & (I_s\gg I_e) \\ 1 & (I_s\ll I_e) \end{array} \right.,
\label{eq:ratio}
\end{equation}
where $I_s$ and $I_e$ are the scattered and self-emission intensity, respectively. The ratio is close to unity if the self-emission intensity is much larger than the scattered intensity, while it is approximated to the ratio of the incident laser if the self-emission is negligible.
The intensities of vertical and horizontal polarization components on the gray markers at $x = -\SI{250}{\micro m}$ are $\sim 10^{3.9}$ and $\sim 10^{4.4}$, respectively. The intensity ratio of vertical to horizontal polarization components is $10^{-0.5} = 0.32$. In this experimental setup, the polarization of the laser is in the $y$ direction. The projection of polarization on the detector coordinate is $\tan^{-1}(\cos \SI{110}{\degree}) \approx \SI{19}{\degree}$ off the horizontal axis and the ratio of vertical to horizontal polarization components is $\cos(\SI{110}{\degree}) \approx 0.34$, thus, the measured intensity ratio is consistent with the drive laser and the observed image corresponds to the scattering of the drive laser, not self-emission. 

\subsection{Pre-pulse dependence}\label{sec:pm}

\begin{figure}[tb]
    \centering
    \includegraphics[clip,width=0.85\hsize]{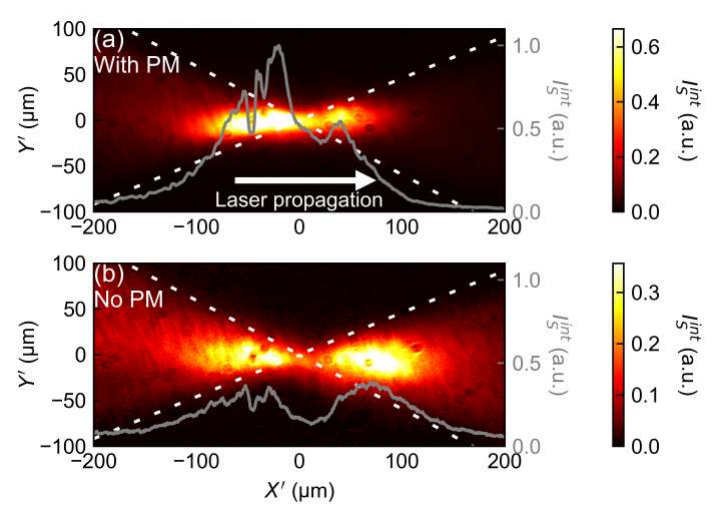}
    \caption{2D images of scattered light using the hydrogen gas target (a) with and (b) without PM.}
    \label{fig:imgpm}
\end{figure}

The candidates to make the scattered intensity lower at the focal spot in Fig.~\ref{fig:imgwp}(b) are the pre-pulse of the drive laser or the time dependence of the density structure due to the ponderomotive force. The pre-pulse heats the electrons mainly near the focal spot because the pre-pulse intensity is larger than the ionization threshold around the focal spot and the laser energy is efficiently absorbed at the focal spot. Because the temperature increases at the focal spot, the electrons diffuse away from the focal spot before the arrival of the main pulse. 
We made another measurement using a plasma mirror (PM) system to control the pre-pulse level \cite{kon22hpl} and observe the effect of the pre-pulse on scattered intensity profile. Figure~\ref{fig:imgpm} shows the images of the scattered light with and without PM using the hydrogen gas target. Since we optically rotate the image by \SI{19}{\degree}, the horizontal and vertical axes on the detector is written as $(X',Y') = (X\cos \SI{19}{\degree} -Y\sin \SI{19}{\degree}, X\sin \SI{19}{\degree} + Y\cos\SI{19}{\degree})$. The color scale is defined so that the upstream intensities in Figs.~\ref{fig:imgpm}(a) and \ref{fig:imgpm}(b) at $X'= \SI{-150}{\micro m}$ are the same. We distinguish the different plasma states by imaging of the scattered light since the spatial distribution is different with and without PM. The focal spot is the brightest in Fig.~\ref{fig:imgpm}(a), while the focal spot is darker than the surroundings in Fig.~\ref{fig:imgpm}(b). This is clear in the gray curves in Figs.~\ref{fig:imgpm}(a) and \ref{fig:imgpm}(b) that show the intensity profile at $Y'=0$. The spatial distribution in Fig.~\ref{fig:imgpm}(b) is similar to that in Fig.~\ref{fig:model}(d). 
The brightest point in Fig.~\ref{fig:imgpm}(a) is located upstream of the focal spot. This can be due to the relativistic self-focusing of the laser pulse \cite{pukhov96prl,makarov17oe}. In our experiment, the total laser power of \SI{500}{TW} is beyond the threshold of the self-focusing; $P_{cr}= m_e^2 c^5 n_c/(e^2 n_e) \approx \SI{700}{GW}$ with a typical electron density of $n_e = \SI{4e19}{cm^{-3}}$, where $m_e$ is the electron mass, $e$ is the elementary charge, and $n_c$ is the critical density \cite{pukhov96prl,makarov17oe}. The scattered light intensity is asymmetry for the focal spot; the upstream is brighter than the downstream. The observed light reflects the electron density profile because of the scattering, the electron density is considered to be lower at the laser irradiation timing in the absence of PM.

\subsection{Non-uniform target}\label{sec:cl}

\begin{figure}[tb]
    \centering
    \includegraphics[clip,width=0.7\hsize]{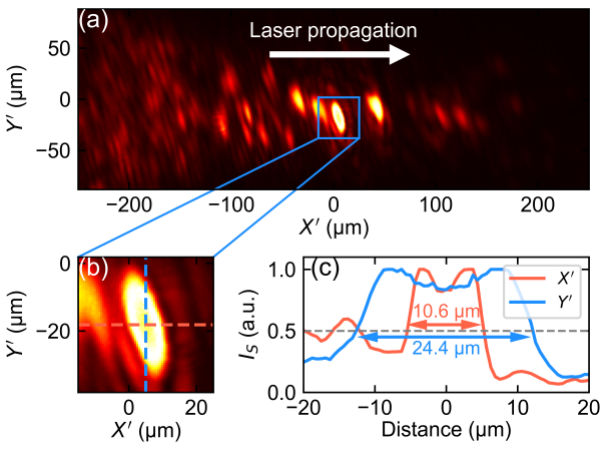}
    \caption{(a) 2D images of scattered light with the hydrogen cluster target. (b) An enlarged view of the cluster close to the focal spot in (a). (c) horizontal and vertical profiles of (b).}
    \label{fig:cls}
\end{figure}

We add a non-uniformity of target using cluster targets \cite{jinno18ppcf,jinno22srep}. Figure~\ref{fig:cls}(a) shows a scattered image with hydrogen clusters. The spatial distribution of scattered light is different from that in Figs.~\ref{fig:imgwp} and \ref{fig:imgpm} due to the different electron density structure. The many grain structures in the image correspond to the clusters interacting with the intense laser beam. According to the Mie scattering measurements, the averaged hydrogen cluster size is $\lesssim \SI{1}{\micro m}$ \cite{jinno18ppcf}, which is less than the spatial resolution of the imaging system. Therefore, the spatial distribution of cluster images shows the spatial resolution of the imaging system. Figure~\ref{fig:cls}(b) is an enlarged view of Fig.~\ref{fig:cls}(a), showing the cluster close to the focal spot at $(X',Y')= (5,-\SI{18}{\micro m})$. The field of view in Fig.~\ref{fig:cls}(b) is indicated by the blue square in Fig.~\ref{fig:cls}(a). The cluster appears to be elliptical because of the tilt of the first lens. It can be circular with a proper alignment. The minor axis of the elliptical shape shows the spatial resolution with the proper alignment. Figure~\ref{fig:cls}(c) shows the horizontal and vertical profiles on the red and blue dotted lines in Fig.~\ref{fig:cls}(b), respectively. The intensity is normalized to the peak count. The horizontal dashed line is the half maximum of the peak intensity. The horizontal and vertical full width at half maximum (FWHM) indicated by arrows are \SI{10.6}{\micro m} and \SI{24.4}{\micro m}, respectively. Note that the red curve in Fig.~\ref{fig:cls}(c) is not the profile of the actual minor axis because the elliptical shape in Fig.~\ref{fig:cls}(b) is tilted in the $X'-Y'$ plane. The spatial resolution of the imaging system is at least $\lesssim\SI{10}{\micro m}$. 

\section{Model calculation} \label{sec:model}
\begin{figure}[tb]
    \centering
    \includegraphics[clip,width=0.7\hsize]{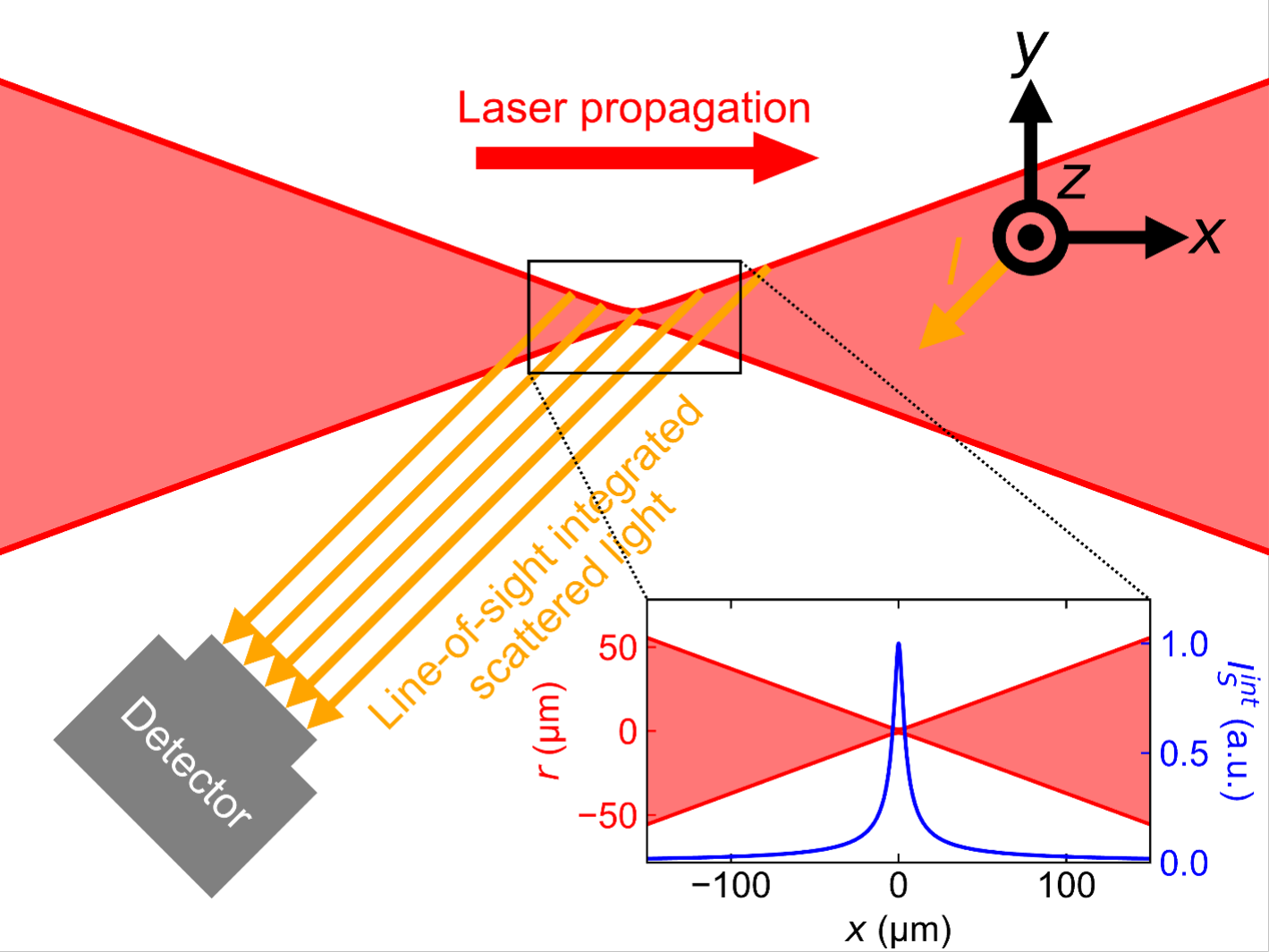}
    \caption{Schematic illustration of the imaging measurement and the line-of-sight integral in Eq. \eqref{eq:sint}. The inset shows the shape of an F/1.35 Gaussian laser beam around the focal spot with the spot size of \SI{2}{\micro m} and the scattered intensity along the $x$ axis assuming a uniform density given by Eq. \eqref{eq:ismodel}.}
    \label{fig:integral}
\end{figure}

The intensity of the scattered light is proportional to the electron density and the intensity of the drive laser \cite{froula11}. The image shows the intensity at the focal spot is weaker than the surroundings. The scattered intensity profile ($I_S^{int}$) is given by the time and line-of-sight integral of
\begin{equation}
    I_S^{int} \propto \iint n_{e}I_L dl dt,
    \label{eq:sint}
\end{equation}
where $I_L$ is the laser intensity. Figure~\ref{fig:integral} shows the integral path of Eq.~\eqref{eq:sint}. The length of the integral path is proportional to the laser beam radius ($r$). Because the laser intensity is proportional to $r^{-2}$ on the laser propagation axis, $I_S^{int} \propto r^{-1}$ assuming a uniform electron density profile without changing in time. The beam radius near the focal spot shows a hyperbolic shape as shown in Fig.~\ref{fig:integral}. The beam radius of a Gaussian beam is written as $r^2 = r_0^2 + x^2/(4F^2)$, where $r_0$ is the focal spot radius. The scattered intensity is proportional to 
\begin{equation}
I_S^{int} \propto \left(r_0^2 + \frac{x^2}{4F^2}\right)^{-0.5}
\label{eq:ismodel}
\end{equation}
and peaks at the focal spot as shown in the blue curve in the inset of Fig.~\ref{fig:integral}. The fact that the experimentally observed intensity at the focal spot is weaker than the surroundings indicates the density profile in the experiment is not uniform in space and the density is lower near the focal spot. 

\begin{figure}[tb]
    \centering
    \includegraphics[clip,width=0.7\hsize]{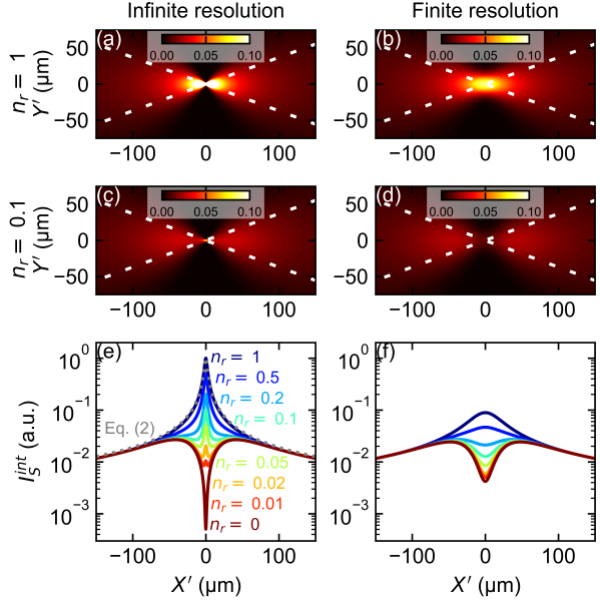}
    \caption{Model calculation results. (a) $n_r=1$ without convolution. (b) $n_r=1$ with convolution. (c) $n_r=0.1$ without convolution. (d) $n_r=0.1$ with convolution. (e) profiles at $y=0$ without convolution. (f) profiles at $y=0$ with convolution.}
    \label{fig:model}
\end{figure}
To include the effect of density non-uniformity, we numerically calculate Eq. \eqref{eq:sint} assuming density profiles.
Figure~\ref{fig:model} shows the line-of-sight integrated profile of scattered light as illustrated in Fig.~\ref{fig:integral}. 
We assume a steady electron density distribution with a cavity in electron density at the focal spot;
\begin{equation}
n_e = n_0 - n_0(1 - n_r) \exp\left[-4\log 2 \frac{x^2+y^2+z^2}{R^2} \right],
\label{eq:nemodel}
\end{equation} 
where $n_0$ is the background density, $n_r$ is the ratio of the density at the focal spot to the background density, and $R$ is FWHM of the cavity at the focal spot. We set $R=\SI{100}{\micro m}$. We effectively include the finite spatial resolution of the imaging system by convolving the obtained distribution of scattered light with a 2D Gaussian function. We assume \SI{20}{\micro m} FWHM of the Gaussian function, which is comparable to the observed width of the focal spot in Fig.~\ref{fig:imgwp}(b). 
Figures~\ref{fig:model}(a) and \ref{fig:model}(b) show the synthetic images in a uniform density plasma ($n_r=1$) without and with the convolution, meaning infinite and finite resolution, respectively. The intensity is normalized to the peak intensity in Fig.~\ref{fig:model}(a). The intensity peaks at the focal spot as indicated by the simple estimation in Eq. \eqref{eq:ismodel}. The highest intensity region is narrow ($\sim \SI{10}{\micro m}$ along the $X'$ axis) because of the tight focusing. Figures~\ref{fig:model}(c) and \ref{fig:model}(d) are the same plots as Figs.~\ref{fig:model}(a) and \ref{fig:model}(b) except for a non-uniform density profile with $n_r=0.1$. Although the density at the focal spot is 10 times smaller than that in the uniform case, the focal spot is brightest in Fig.~\ref{fig:model}(c) because the intensity peaks at the focal spot. The surroundings are brighter than that in the uniform density case in Fig.~\ref{fig:model}(a). When we include the effect of resolution by calculating the convolution, the relatively lower intensity at the focal spot affects the spatial distribution and the focal spot becomes dark. 
We compare the spatial distribution on $Y'=0$ in Figs.~\ref{fig:model}(e) and \ref{fig:model}(f). The different color corresponds to the different $n_r$. The gray dotted curve in Fig.~\ref{fig:model}(e) is the profile with the simple model in Eq.~\eqref{eq:ismodel}. This model agrees well with the $n_r=1$ profile. The intensity profile in Fig.~\ref{fig:model}(e) shows that the intensity at the focal spot is not so dark as the surroundings with $n_r\gtrsim 0.02$, however, taking account of the resolution in Fig.~\ref{fig:model}(f), the focal spot becomes dark with $n_r\lesssim 0.1$. The observed image in Fig.~\ref{fig:imgwp}(b) shows that the scattered intensity at the focal spot is $10^{4.2}\approx 16000$ and that at the surroundings is $10^{4.6}\approx40000$, i.e., the intensity at the focal spot is $\gtrsim 2$ times less than that at the surroundings. The model calculation results including finite resolution effect in Fig.~\ref{fig:model}(f) shows that $\gtrsim 2$ times less intensity at the focal spot is achieved when $n_r\lesssim 0.1$ [red, orange, yellow, and green curves in Fig.~\ref{fig:model}(f)]. Therefore, the electron density at the focal spot can be less than 10\% of the initial target density to explain the observed intensity ratio.

The intensity at the focal spot with PM in Fig.~\ref{fig:imgpm}(a) is not darker than that at the surroundings. The model calculation results with finite resolution show that the drop of scattered intensity at the focal spot is not observed when $n_e\gtrsim 0.2$ (blue curves in Fig.~\ref{fig:model}(f)), therefore, the electron density is more than 20\% of the initial target density. Because the focal spot density without PM is less than 10\%, the electron density interacting with the main pulse at the focal spot increases more than a factor of two with PM.

\section{Numerical simulation} \label{sec:sim}

We performed 2D particle-in-cell (PIC) simulations with the Smilei open source code \cite{derouillat18cpc} to take into account the time dependence of the electron density structure in the scattered images. A laser pulse linearly polarized in the $y$ direction propagates along the $x$ direction. The laser parameters are the same as the experimental condition (see Sec. \ref{sec:exp}). The focal spot of the laser beam is located at the center of the simulation domain. The peak of the laser pulse arrives at the left boundary of the simulation domain at $3\tau_L = \SI{120}{fs}$. The simulation domain is initially filled with a pre-ionized hydrogen plasma with the background density of $n_{0} = 0.024 n_c \approx \SI{4e19}{cm^{-3}}$. We consider two density structures: uniform [$n_r=1$ in Eq.~\eqref{eq:nemodel}] and non-uniform plasmas ($n_r=0.1$). We use the realistic mass ratio of proton to electron; $m_i/m_e=1836$. The number of macroparticles is $2.7\times 10^{9}$. The length of the simulation domain is $(L_x,L_y)=(512 \lambda_L, 256 \lambda_L) \approx (420, \SI{210}{\micro m})$ divided into $(N_x,N_y)=(16384, 8192)$ cells. The total computational time is $T \approx 48 \tau_L = \SI{1.9}{ps}$, which is larger than $L_x/c \approx 35 \tau_L = \SI{1.4}{ps}$ so that the drive laser pulse passes through the transverse simulation domain. We use an open boundary condition. 

\begin{figure}[tb]
    \centering
    \includegraphics[clip,width=0.7\hsize]{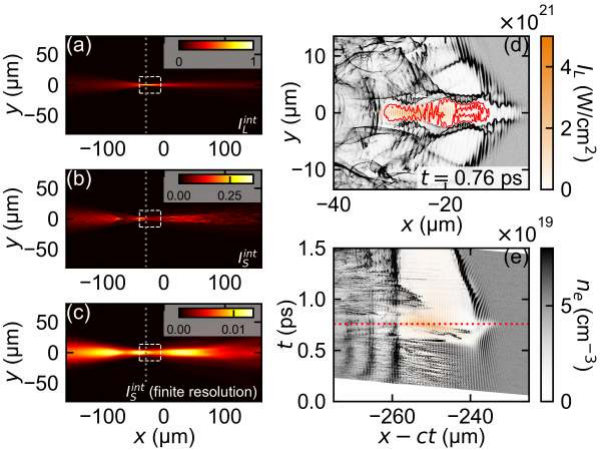}
    \caption{2D PIC simulation with $n_r=1$. The time-integrated maps of (a) laser intensity and (b) scattered intensity. We take the finite resolution effect into account in (c). (d) The snapshot and (e) time evolution of electron density and laser intensity.}
    \label{fig:simu}
\end{figure}

Figure~\ref{fig:simu} shows the results with a uniform density plasma; $n_r = 1$. The time-integrated laser intensity map in Fig.~\ref{fig:simu}(a) is calculated with $I_L^{int} = \int I_L dt$. The color is normalized to the peak intensity. The peak is upstream of the nominal focal spot at the origin due to the relativistic self-focusing \cite{pukhov96prl,makarov17oe}. The peak position at $x= -\SI{29}{\micro m}$ is indicated by the vertical dotted line. 
We obtain synthetic scattered images in Fig.~\ref{fig:simu}(b) by calculating $I_{S}^{int} = \int n_e I_L dt$. The laser intensity is proportional to $I_L \propto B_z^2$. The color is normalized to the peak intensity with a uniform plasma assuming no temporal change of density. The normalized scattered intensity at the focal spot is less than unity because the electron density is lower than the target density when the laser pulse arrives at the focal spot. As considered in Sec.~\ref{sec:pol}, the finite resolution of the measurement system is effectively included in Fig.~\ref{fig:simu}(c) by calculating the convolution with a 2D Gaussian function with a \SI{20}{\micro m} FWHM. The focal spot is not as bright as the expectation in Fig.~\ref{fig:model}(b). 
Figure~\ref{fig:simu}(d) shows a snapshot of electron density (gray) and laser intensity (orange) at $t=\SI{0.76}{ps}$. The plotted area in Fig.~\ref{fig:simu}(d) is indicated by the dashed squares in Figs.~\ref{fig:simu}(a)--\ref{fig:simu}(c). There is a cavity in the electron density at the laser pulse. The red contour shows $n_e = 0.1 n_0$, which corresponds to the upper limit to obtain a weak intensity at the focal spot according to the analytic model in Sec.~\ref{sec:pol}. Since most of the laser pulse is located inside the contour, few electrons interact with the intense laser pulse. The electron density varies in space due to the strong ponderomotive force and this is the reason why the synthetic scattered image in Fig.~\ref{fig:simu}(c) does not have a clear peak at the focal spot. 
The time evolution of electron density and laser intensity at $y=0$ is shown in Fig.~\ref{fig:simu}(e). The color is the same as in Fig.~\ref{fig:simu}(d). The horizontal axis is the frame moving in the $+x$ direction at the speed of light. The horizontal dotted line shows the timing of Fig.~\ref{fig:simu}(d). The less-density region forms at $\gtrsim \SI{0.5}{ps}$ and remains after that. The averaged electron density at the upstream of focal spot before focusing ($t\lesssim \SI{0.5}{ps}$) is close to the background density.  

\begin{figure}[tb]
    \centering
    \includegraphics[clip,width=0.7\hsize]{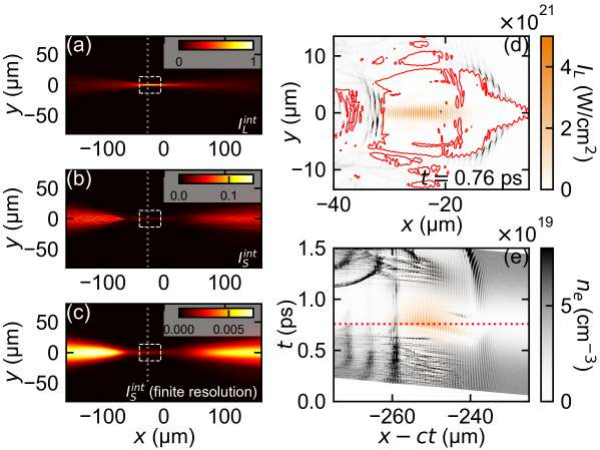}
    \caption{2D PIC simulation with $n_r=0.1$. The time-integrated maps of (a) laser intensity and (b) scattered intensity. We take the finite resolution effect into account in (c). (d) The snapshot and (e) time evolution of electron density and laser intensity.}
    \label{fig:simn}
\end{figure}

Figure~\ref{fig:simn} is the same plot as Fig.~\ref{fig:simu} but a non-uniform density plasma; $n_r = 0.1$. The focal spot shown in Fig.~\ref{fig:simn}(a) is at $x= -\SI{26}{\micro m}$ and is \SI{3}{\micro m} closer to the origin or the nominal focal spot than that in Fig.~\ref{fig:simu}(a). The synthetic scattered images in Figs.~\ref{fig:simn}(b) and \ref{fig:simn}(c) show less intensity at the focal spot, which is similar to the model in Fig.~\ref{fig:model}(d). At the interaction region in Fig.~\ref{fig:simn}(d), the electron density at the laser pulse is less than $0.1 n_0$. As shown in Figs.~\ref{fig:simu}(d) and \ref{fig:simu}(e), the scattered intensity at the focal spot is mainly explained by the interaction between the tail of the laser pulse and the dense region at $x-ct= -\SI{260}{\micro m}$ that corresponds to a wakefield.

\section{Discussion and summary}

We confirmed the observed signal is the scattered light of intense laser itself in Fig.~\ref{fig:imgwp}(b) because the observed signal has a polarization consistent with the drive laser beam. The scattering occurs while the incident laser interacts with the electrons. The pulse duration of the drive laser beam is \SI{40}{fs} and thus the interaction time for the electrons at rest on average is \SI{40}{fs}. Therefore, the temporal resolution of the images is determined by the pulse duration of the drive laser beam. The imaging of scattered intense lasers can provide the spatial distribution of electrons at the time interacting with the intense laser beam with a high temporal resolution. 

Both experimental and numerical results indicate that the spatial structure of the scattered images can be attributed to both the laser pre-pulse and the spatiotemporal change of the density; the global structure of the images is mainly determined by the global density structure generated by the pre-pulse, while the local structure at the focal spot is strongly dependent on the temporal variation of the electron density due to the strong ponderomotive force. Observed structures also depends on the spatial resolution of the measurement system. With a high spatial resolution, the fine structure at the focal spot should appear, as shown in Fig.~\ref{fig:simu}(b). The high spatial resolution is achieved using larger optical lenses with smaller focal lengths.
The detailed structures and the elementary behavior of electrons are observed with the gas-density target, which is uniform and has no clear interaction point unlike a solid density target, e.g., the cluster target in Fig.~\ref{fig:cls}.

The results of the numerical simulations show that the electron density upstream of the focusing laser beam is close to the background density. This means that the upstream scattered intensity can be explained by the simple model in Eq.~\eqref{eq:ismodel} because the electron density is constant in time and space during the laser pulse. When we measure the far upstream together with the focal spot and downstream, we can estimate the absolute value of the electron density from the upstream. The absolute calibration of the target density upstream is necessary for this estimation. The field of view should be greater than the current one to estimate the absolute density. 

As shown in Fig.~\ref{fig:imgpm}(a), the upstream intensity is higher than the downstream one in the presence of PM. The result with PM is closer to the situation in Fig.~\ref{fig:simu} because of less pre-pulse. Looking at the time evolution of electron density in Fig.~\ref{fig:simu}(e), the electron density becomes 0 at the laser focusing time due to the strong ponderomotive force and the electron evacuated structure remains after that. The less density structure after the focusing time or at the downstream results in less intensity at the downstream. In order to confirm this scenario, we need time evolution of electron density structure, e.g., imaging using a streak camera.

There is a wavelength dependence because of the Doppler shift in wavelength excited by the electron bulk flow due to strong ponderomotive forces. Selecting the red-shifted wavelength, we can observe the density of electrons moving in the direction of the laser propagation.

The upper limit of the measurable density is the critical density because the laser beam should propagate in the plasma as a probe. The lower limit is determined by the dynamic range of the detector. In the current setup, we use a CCD camera with 16-bit analog-to-digital conversion. Therefore, the observed count is ranging from 0 to 65535. The typical readout noise is 10 count. If the brightest part of the image is at the maximum count without saturation, the minimum count we can observe is 10, which is three orders of magnitude less than the brightest part. As shown in Eq. \eqref{eq:sint}, the signal intensity is proportional to the product of local electron density and laser intensity. The observed space is $\pm \SI{200}{\micro m}$ from the focal spot in the current setup. At the edge of the field of view, the line-of-sight integrated laser intensity is two orders of magnitude smaller than that at the focal spot. Assuming that the brightest part is at the edge of the field of view, the lowest measurable density at the focal spot is five orders of magnitude less than the background density.

In summary, we reported the experimental results on the scattered light in intense laser beams. The scattered intensity is weak at the focal spot of the laser beam. With a higher pulse contrast using the plasma mirror system, the scattered intensity becomes high at the focal spot. Therefore, the pre-pulse destroys the target before the arrival of the drive laser pulse. The elementary process of electron response is visualized with the imaging using the gas-density target. The imaging of scattered intense laser beam can be used for the density structure measurement at the laser irradiation timing with a temporal resolution comparable to the pulse duration. The imaging diagnostic presented in this paper is easy to construct because the scattered light is much brighter than other signals such as self-emission. It is possible to observe, for instance, the Weibel filaments in sub-relativistic collisionless shock experiments \cite{kuramitsu23pop} and the structures of wakefields in electron acceleration experiments \cite{nakanii15prab}.

In this paper, we observed the spatial distribution of the scattered light. In addition to this, the spectrum of the scattered light should have more detailed information on the interaction between intense laser and plasma. While the scattering in a weak laser beam that has little effect to change the plasma state is well-understood \cite{froula11}, intense lasers drastically change the plasma state. The scattering from time-varying plasmas is highly complicated even with a weak laser beam \cite{sakai20pop,sakai23pop}. Therefore, there are no theoretical frameworks to quantitatively explain the scattered spectrum in intense laser. We are developing a spectroscopy system to experimentally understand the scattered spectrum in intense laser beams and to measure the plasma state other than the electron density during the interaction of intense laser and plasma. 

\bmsection*{Acknowledgments}
The J-KAREN-P experiment was performed under the Shared Use Program of QST Facilities (2022A-K08).
This work was the result of using research equipment shared in MEXT Project for promoting public utilization of advanced research infrastructure (Program for advanced research equipment platforms) Grant Number JPMXS0450300221.
The numerical simulations were performed using the supercomputer of ACCMS, Kyoto University through the HPCI System Research Project (Project ID: hp240035).
This work was supported by JSPS KAKENHI (25H00622, 24K17029, 24H01624, 23K20038, JSCCA20230003, 22H01195, 22K14020, 21J20499, 20KK0064, 19K21865, 19H00668, JSBP120203206, and JSCCA2019002); 
by the grant of Joint Research by NINS (OML022405, OML012508, OML012509); 
by the NINS program of Promoting Research by Networking among Institutions (01422301, 01412302); 
by the NIFS Collaboration Research Program (NIFS24KIIQ010);
by i-SPES collaborative research project. 

\bibliography{ctpp}

\end{document}